\documentclass[aps,prl,twocolumn,a4paper,10pt]{revtex4-1}

\usepackage[T1]{fontenc}		
\usepackage[english]{babel}		
\usepackage[latin1]{inputenc}	
\usepackage{times}

\usepackage{amsmath}			
\usepackage{amssymb}			
\usepackage{bbm}				

\usepackage[colorlinks=true, a4paper=true, pdfstartview=FitV,linkcolor=blue, citecolor=blue, urlcolor=blue]{hyperref}

\usepackage{graphicx}			
\usepackage{graphics}			
\DeclareGraphicsExtensions{.pdf}
\usepackage{color}		

\newcommand{\bea}{\begin{eqnarray}}
\newcommand{\eea}{\end{eqnarray}}

\newcommand{\bk}{\mathbf{k}}

\newcommand{\bsigma} {\boldsymbol{\sigma}}

\newcommand{\bd}{\mathbf{d}}

\begin{document}

\title{Knotted Non-Hermitian Metals}

\author{Johan~Carlstr\"om$^{1}$, Marcus St{\aa}lhammar$^{1}$, Jan Carl Budich$^{2}$ and Emil~J. Bergholtz$^{1}$}
\affiliation{$^1$Department of Physics, Stockholm University, AlbaNova University Center, 106 91 Stockholm, Sweden\\
$^2$Institute of Theoretical Physics, Technische Universit\"{a}t Dresden, 01062 Dresden, Germany}

\date{\today}

\begin{abstract}
We report on the occurrence of knotted metallic band structures as stable topological phases in non-Hermitian (NH) systems. These knotted NH metals are characterized by open Fermi surfaces, known in mathematics as Seifert surfaces, that are bounded by knotted lines of exceptional points. Quite remarkably, and in contrast to the situation in Hermitian systems, no fine tuning or symmetries are required in order to stabilize these exotic phases of matter. By explicit construction, we derive microscopic tight-binding models hosting knotted NH metals with strictly short-ranged hopping, and investigate the stability of their topological properties against perturbations. Building up on recently developed experimental techniques for the realization of NH band structures, we discuss how the proposed models may be experimentally implemented in photonic systems.   
\end{abstract}
\maketitle


{\textit{Introduction.---}} Revealing the topological properties of Bloch bands has revolutionized the theory of solids: Fascinating new forms of quantum matter such as topological insulators \cite{hasankane,qizhang} and Weyl semimetals \cite{weylreview} have been discovered and theoretically described in terms of topological invariants that measure how the phase of the Bloch functions twists in reciprocal space \cite{jantopreview}. Recently, the experimental discovery of topological phases in various dissipative systems subject to gain and loss \cite{NHarc,wiemannkremerplotniklumernoltemakrissegevrechtsmanszameit,EPringExp,NHtransition,NHexp2,NHlaser,asymhop1,asymhop2}  has triggered the urgent quest for generalizing this topological band theory to non-Hermitian (NH) systems \cite{reviewTorres,kunstedvardssonetc,gong,koziifu,jan,yoshidapeterskawakmi,xiong,
yaowang,NHchern, EPrings,carlstroembergholtz,nodal5,leethomale,lee,lieu,schomerus,shenzhenfu,yuce,malzard,esaki,lieu2, kawabatahisgashikawagongashidaueda,eringchiral,esurfptsym,esurfptpphsym,exhopflinkpert,disorderRing,disorderlinesribbons,molina,yaosongwang}. 
While a comprehensive understanding of the role of topology in NH systems is still lacking, several crucial differences to the Hermitian realm have already been established. Prominent examples of these include a modified relation between bulk topological invariants and protected surface states (bulk boundary correspondence) \cite{kunstedvardssonetc,yaowang,leethomale} as well as qualitative changes in the topological stability of nodal surfaces \cite{carlstroembergholtz,koziifu,EPrings,jan}. Regarding the latter, extending the notion of band touching points to that of exceptional points (EPs) in NH systems, leads to the occurrence of nodal band structures in lower spatial dimensions \cite{BerryDeg,carlstroembergholtz}, where conventional band touching points would be unstable, or dependent on symmetries.
Thus, it should be stressed that while there exists a number of works on nodal topology in the form of lines \cite{NodalLineCentroSym} chains, \cite{Nodal-Chain} and even knots \cite{nodalknotsemimetals}, which is the closest Hermitian counterpart to this work, the inclusion of non-Hermitian effects leads to a profoundly different scenario, with nodal structures depending only on topology, and not on any symmetries of the system.   

\begin{figure}[t]
\centering
\includegraphics[width=\columnwidth]{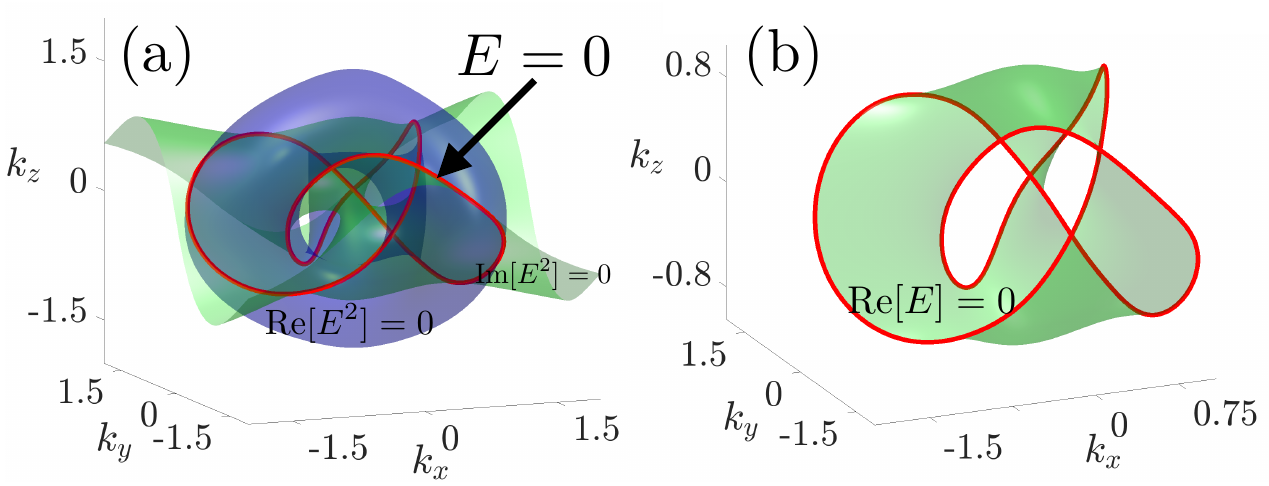}
\caption{\label{fig:one}Illustration of a knotted nodal band structure obtained from a tight-binding model with a hopping range of two lattice constants defined by Eqs. (\ref{dvector},\ref{f_explicit}) with $\Lambda=-20$. The solid blue line is a trefoil knot corresponding to the exceptional lines of the model that correspond to the intersection of the surfaces $\text{Re}[E^2]=0$ and $\text{Im}[E^2]=0$ which are marked red and green respectively in (a). (b) Shows the same exceptional knot (blue) and its concomitant open $\text{Re}[E]=0$ Fermi surface (green). This trefoil knot is also called the $(3,2)$ torus knot and is the most elementary representative of the general family of $(p,q)$ exceptional torus knots constructed in this work.}
\end{figure}

Drawing intuition from this fundamental observation, here we show and exemplify how topologically stable nodal lines can naturally form knots in reciprocal space in three-dimensional (3D) NH systems (see Fig.~\ref{fig:one} for an illustration), thus introducing a genuinely NH class of metallic topological phases. Remarkably, these knotted lines of EPs necessarily form the boundaries of open Fermi surfaces, known in mathematics as Seifert surfaces, the topological properties of which are closely related to a knot invariant characterizing the nodal line. In this sense, knotted lines of EPs result in knotted non-Hermitian metals rather than semimetals. It is easy to see that this intriguing phenomenon has no counterpart in generic Hermitian systems. This is because knotted closed lines can be topologically nontrivial only when embedded in a 3D space, where band touchings in Hermitian systems without symmetries or fine-tuning only occur at isolated points.

Below, we present principles for the construction of generic non-Hermitian two-band models with knotted lines of exceptional points.
This general approach is illustrated by the explicit construction of microscopic tight-binding models with strictly short-ranged hopping that realize paradigmatic examples of knots such as the trefoil knot shown in Fig.~\ref{fig:one}. For a Hopf link representing a precursor of a knot, our construction simplifies to yield a nearest neighbor tight-binding model. Furthermore, we investigate the stability of the knotted nodal lines against perturbations, and discuss platforms for the experimental realization of knotted non-Hermitian metals.

{\it Knotted exceptional lines and open Fermi surfaces.---} An interaction free system with two bands is fully characterized in reciprocal space by its Bloch Hamiltonian which may be written as
\bea
H(\mathbf k)=\bd_R(\mathbf k) \cdot \bsigma + i\bd_I(\mathbf k)\cdot \bsigma,
\eea
where $\mathbf k$ denotes the lattice momentum, $\bsigma=(\sigma_x,\sigma_y,\sigma_z)$, and $\bd_R,\bd_I\in \mathbb{R}^3$. Nodal points then correspond to coalescence of the eigenvalues
\bea
E^2 (\mathbf k)= \bd_R^2(\mathbf k)-\bd_I^2(\mathbf k)+2i \bd_R(\mathbf k)\cdot \bd_I(\mathbf k)\label{eigenValues}
\eea
which is satisfied when both
\begin{eqnarray}
\text{Re}[E^2]=\bd_R^2-\bd_I^2=0\label{realsquared}
\end{eqnarray}
and 
\begin{eqnarray}
\text{Im}[E^2]=2\bd_R\cdot\bd_I=0.\label{imaginarysquared}
\end{eqnarray}
Except for the trivial case $\bd_R(\mathbf k) =\bd_I(\mathbf k) =0$, these solutions correspond to exceptional points where the Hamiltonian becomes defective, i.e. non-diagonalizable with only a single eigenvector associated with the two-fold degenerate eigenvalue \cite{Heiss}.

In three dimensions, the solutions to Eq. (\ref{realsquared}) and Eq. (\ref{imaginarysquared}) each describe closed surfaces, with intersections in the form of closed lines. These are in turn connected by open Fermi surfaces \cite{carlstroembergholtz}, as follows from the eigenvalue equation (\ref{eigenValues}): Requiring $\text{Re}[E]=0$, we find
 \bea
\text{Im} [E^2]=0,\;\text{Re}[E^2] \le 0, \label{fermiSurface}\;
 \eea
 implying that the Fermi surface $ \text{Re}\;[E]=0$ is a subset of the closed surface $\text{Im }[E^2]=0$, and is as such generally open. Similarly, we can define an open i-Fermi surface by requiring  $\text{Im}[E]=0$ which leads to the conditions
 \bea
\text{Im}[E^2]=0,\;\text{Re}[E^2]\ge 0.\label{ifermiSurface}\;
 \eea
 The existence of a finite Fermi surface, described by Eq. (\ref{fermiSurface}), indicates that these systems are metals, in contrast to  Hermitian line-node semimetals that exhibit a vanishing density of states at the Fermi level. 

If particle-hole symmetry is broken, then the exceptional line is shifted to a finite energy. In this scenario, Eqs. (\ref{fermiSurface}, \ref{ifermiSurface}) do instead describe the inter-band gap, rather than a Fermi surface. 



The insight that exceptional lines occur at the intersections of orientable surfaces forming the solutions of  Eqs.~(\ref{realsquared},\ref{imaginarysquared}) provides a natural starting point for generating non-Hermitian models with topologically non-trivial nodal lines: Consider two surfaces $s_1$ and $s_2$ in momentum space with a topologically nontrivial intersection. Next, assume two real and continuous scalar functions $f_1$ and $f_2$ that vanish on these surfaces, i.e.

\bea
f_i(\bk\in s_i)=0,~i=1,2. \label{f_i} 
\eea
Then we may encode the topology of the intersection of $s_1$ and $s_2$ into our model Hamiltonian by taking
\bea
\!\!\bd_R(\mathbf k)\!=\!\big[f_1(\mathbf k)\!-\!\Lambda,\Lambda,0\big],\;
\bd_I(\mathbf k)\!=\!\big[0,f_2(\mathbf k),\!\sqrt{2}\Lambda\big]\label{dvector}
\eea
which, using Eqs. (\ref{realsquared},\ref{imaginarysquared}),  gives 
\bea
\text{Re}[E^2]=f_1^2-f_2^2-2f_1 \Lambda,\;\;\;\label{nodalEquation}
\text{Im} [E^2]=2f_2\Lambda.
\eea
In the limit of $\Lambda\to\pm\infty$, we obtain solutions to Eq. (\ref{realsquared},\ref{imaginarysquared}) exactly at the intersection $s_1\cap s_2$, 
where changing the sign of $\Lambda$ results in an interchange of the Fermi and i-Fermi surfaces respectivly.  
Finite values of $|\Lambda|$ lead to deformation of the surfaces $\text{Re }[E^2]=0,\;\text{Im }[E^2]=0$, according to Eq. (\ref{nodalEquation}). Thus, given a sufficiently large magnitude of $\Lambda$, the nodes of the model Eq. (\ref{dvector}) will inherit the topology of $s_1\cap s_2$. 
Furthermore, since the nodes occur at finite $\bd_R$, $\bd_I$, they are necessarily exceptional, i.e. they correspond to defective lines of the Hamiltonian. 

To generate the functions $f_1, f_2$ such that the intersection of $s_1$ and $s_2$ obtains a nontrivial topology, we note that a torus knot or link may be described as the solution to a complex algebraic equation on the unit three-sphere. Specifically, consider a pair of complex scalars $(Z_0,Z_1)\in \mathbb{C}^2$ and a corresponding unit three-sphere $S^3$ given by 
\bea
|Z_0|^2+|Z_1|^2=1. \label{S3}
\eea
The $(p,q)$ torus knot or link is then described by the solution to 
\bea
g(Z_0,Z_1)=Z_0^p+Z_1^q=0 \label{complexSurfEq}
\eea
on $S^3$, i.e. subject to the constraint in Eq. (\ref{S3}). 
To generate knots in the band structure, we then map the momentum space onto the three-sphere so that, $Z_0,\;Z_1$ obtain an explicit $\bk-$dependence that satisfies Eq. (\ref{S3}). 
While there exist many ways to achieve this, here we follow the construction outlined in Ref. \onlinecite{sutcliffe} (see also supplementary material \cite{supmat}). 
It should be stressed that even though some elements are set to zero in Eq. (\ref{dvector}), the construction outlined above does not in principle require any symmetry of the system, and remains intact under arbitrary perturbations up to a certain magnitude. 

Having connected $g(Z_0,Z_1)$ to the momentum space we note, following the example of \cite{nodalknotsemimetals}, that it can be parameterized in terms of two real scalar functions by defining 
\bea
f_1(\bk)+if_2(\bk)=g(Z_0(\bk),Z_1(\bk)). \label{rationalMapAnsatz}
\eea
Then, according to Eq. (\ref{complexSurfEq}), the functions $f_1, f_2$ satisfy Eq. (\ref{f_i}), with $s_1, s_2$ being orientable surfaces such that their intersection $s_1\cap s_2$ describes a $(p,q)$ knot or link. Furthermore, according to Eqs. (\ref{fermiSurface}) and (\ref{nodalEquation}) the Fermi surface is now a subset of $s_2$.

\begin{figure*}[!htb]
 \hbox to \linewidth{ \hss
\includegraphics[width=\linewidth]{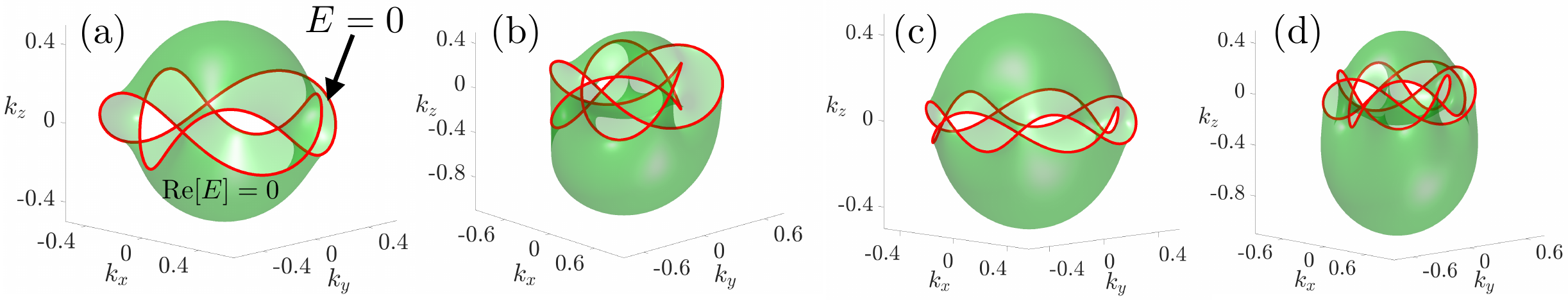}
 \hss}
\caption{
Nodal knots and their concomitant open Fermi surfaces generated from the construction (\ref{dvector}) with $\Lambda=40$ and the ansatz (\ref{rationalMapAnsatz}). The knotted exceptional lines (blue solid line) terminate the Fermi surfaces (green) which are as such open---a unique bulk feature of non-Hermitian systems that is a direct consequence of the non-analyticity of the exceptional points. The nodal topologies displayed here correspond to a choice of coprime integers $(p,q)$ of $(5,2)$, $(5,3)$, $(7,2)$ and $(7,3)$ in the figures (a), (b), (c) and (d) respectively.  
}
\label{knots}
\end{figure*}

A few examples of nodal topologies generated from this procedure are displayed in Fig. \ref{knots}. Here, for the exceptional lines to form torus knots, we stress that $(p,q)$ need to be chosen as coprime. Otherwise, Eq. (\ref{rationalMapAnsatz}) produces links rather than single knots.  
Notably,  it is not possible to uniquely determine the topology of the Fermi surface solely from the knot, though some conclusions may be drawn: The genus of a knot is defined as the minimal genus of a Seifert surface whose boundary it forms \cite{seifert}. Thus, given the genus of a torus knot
\bea
G_{\text{knot}}=(p-1)(q-1)/2,\label{knotGenus}
\eea
we obtain a lower bound for the genus of the Fermi surface, or indeed the i-Fermi surface
\bea
G_{\text{Fermi}}\ge(p-1)(q-1)/2 . \label{fermiGenus}
\eea
This result corresponds to the intuition that the topology of the Fermi surface (i-Fermi surface) can acquire additional handles, hence increasing its genus, without altering its open boundary given by the exceptional knot. For a more recent source on knot theory and Seifert surfaces, see \cite{Adams}.


{\it Knotted tight-binding models.---} To realize exceptional knots in explicit lattice models using the methods described above, it is clear that the mapping from $\bk-$space now should connect the three-sphere with the Brillouin zone rather than $\mathbb{R}^3$. 
However, a key point to note at this stage is that knots may in principle be parameterized equally well on highly deformed three-spheres as long as their topology remains intact.
This fact may be exploited to obtain a mapping where $Z_0,\;Z_1$ are linear functions of short-range hoppings, which in turn translates to a simpler tight-binding model. Here, we thus use a construction of the form
\bea\nonumber
 Z_0(\bk)&=&\sin k_x+i\sin k_y ,\\
Z_1(\bk)&=&2\sum_\alpha \cos k_\alpha-5+i\sin k_z ,\label{M_lattice}
\eea
where $f_1,\;f_2$ are as before given by Eqs. (\ref{complexSurfEq}) and (\ref{rationalMapAnsatz}).

As an example, we consider the trefoil knot corresponding to $(p,q)=(3,2)$, which represents the simplest torus knot and is described by
\bea
f_1(\bk)+if_2(\bk)=Z_0^3(\bk)+Z_1^2(\bk).
\eea
The highest order terms, and thus the most long-ranged dispersion results from the third order terms $\sim \sin^3$ in $Z_0^3$. Introducing the approximation 
$\sin^2 x\approx 2-2\cos x$, which is correct up to $O(x^4)$, the resulting theory contains terms up to quadratic order, implying that the range of the hopping is now two unit cells. 
The explicit form of $f_1,\; f_2$ is then given by
\bea\nonumber
f_1\!&=&\!30.5\!-\!20\sum_{\alpha} \!\cos k_\alpha+2\sum_\alpha \!\cos 2k_\alpha  
+\frac{1}{2}\!\cos 2k_z \;\;\;\;
\\\nonumber
\!&+&\!4\!\sum_{\alpha>\beta}\! \cos(k_\alpha\!\pm \!k_\beta)\!+\!3\sin(k_x\!\pm\! k_y)\!-\!\sin 2k_x
\!-\!4\sin k_x
\\\nonumber
f_2&=&4\sin k_y-10\sin k_z+\sin 2k_y+2\sin 2k_z
\\
&\mp& 3\sin(k_x\pm k_y) \! \pm \! 2\sin(k_y\pm k_z)\! \pm\! 2\sin(k_x\pm k_z), \!\!\!\label{f_explicit}
\eea
where $\pm$ denotes a summation over $(-1,1)$. Albeit somewhat complex in structure it is worth noting that the hopping amplitudes decay rapidly with distance and are smallest for the longest range hopping of two lattice constants. 
The corresponding tight-binding model, which results from inserting Eq. (\ref{f_explicit}) into Eq.  (\ref{dvector}) with $\Lambda=-20$, is displayed in Fig. \ref{fig:one}.

Proceeding to knots of higher genus, the complexity of the lattice models naturally increases. To obtain the $(5,2)$ and $(5,3)$ nodal structures displayed in Fig. (\ref{knots}, a and b), we must include terms $\sim Z_0^5$. Exploiting trigonometric identities and approximations as above, it is possible to arrive at a description where the range of hopping is up to three unit cells. 

\begin{figure}[t]
\centering
\includegraphics[width=\columnwidth]{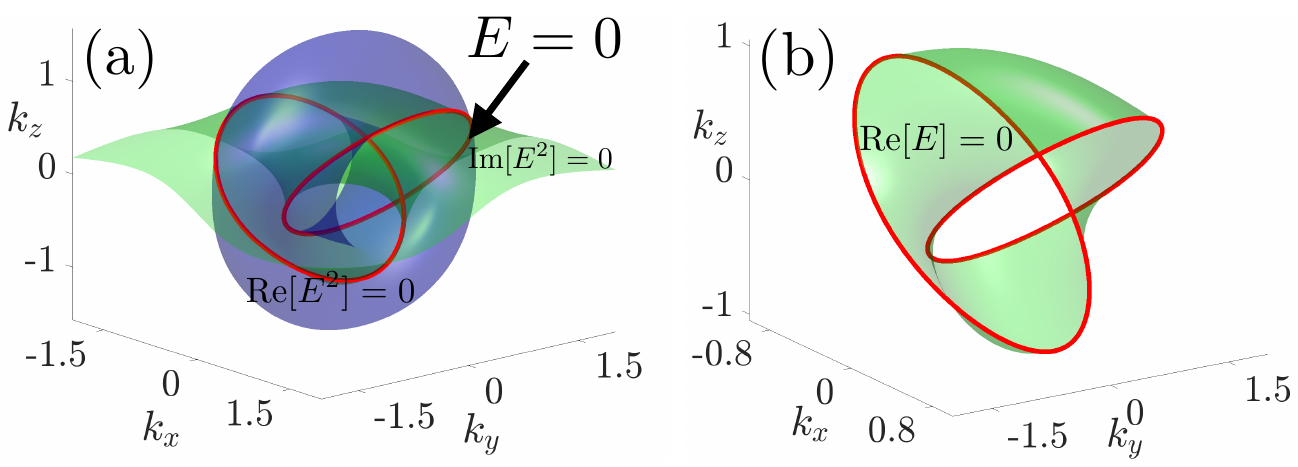}
\caption{Exceptional links (a) and Fermi surface (b) obtained from the nearest neighbor tight-binding model defined by Eq. (\ref{linkHamiltonian}) with $(p,q)=(2,2)$. This model is realizable with current experimental techniques and also features a remarkably high stability towards perturbations.}
\label{easyLink}
\end{figure}

We note that if $(p,q)$ are both even, there is a much simpler way to encode them into a nodal structure than the generic construction given by Eqs. (\ref{dvector},\ref{complexSurfEq}), by instead taking
\bea
H=Z_0^{p/2}(\bk)\sigma_i +Z_1^{q/2}(\bk)\sigma_j,\;i\not=j,\label{linkHamiltonian}
\eea
which gives exceptional nodes described by
\bea
E^2=Z_0^{p}(\bk) +Z_1^{q}(\bk)=0,
\eea
immediately reproducing Eq.  (\ref{complexSurfEq}). A key advantage of using this additional structure is that the resulting tight-binding models become both simpler and shorter ranged. For example, the Hopf link given by $(p,q)=(2,2)$ can be obtained already from nearest neighbor hopping by using the map (\ref{M_lattice}), see Fig. \ref{easyLink}. 
However, to obtain knots rather than links, we stress that it is necessary that $(p,q)$ are coprime integers.  

{\it Stability.---}
We emphasize that nodal lines, knots or links in non-Hermitian systems differ fundamentally from their Hermitian counterparts in the sense that their realization or stability does not depend on the existence of symmetries or fine-tuning. This circumstance is manifested in the fact that band touching points are described by only two equations 
(\ref{realsquared},\ref{imaginarysquared}) so that they generically become line-like in three spatial dimensions.
Correspondingly, the nodal knots that we discuss here are stable in the sense that their topology cannot be altered by any deformation until the point where reconnections or crossings of exceptional lines occur, or where their size is shrunk to zero. 
Thus, introducing perturbations to the Hamiltonian of the from
\bea
H(\bk)\to H(\bk)+\sum_{i=x,y,z}\delta_i \sigma_i,\;\delta_i\in \mathbb{R}
\eea
we obtain a finite range of perturbations $\delta_i$ in which the nodal topology remains intact. We note that similar arguments hold for anti-Hermitian perturbations, upon multiplying the entire Hamiltonian by the imaginary $i$.

On a quantitative note, for the trefoil knot described by Eq. (\ref{f_explicit}) with $\Lambda=-20$, we find that the critical magnitudes $\delta_i^c$ of perturbations are anisotropic, but all fall into the range $0.4<|\delta_i^c|<0.7$.
Comparing to (\ref{f_explicit}), we infer that this corresponds to $4-7\%$ of the largest nearest neighbor hopping integral. 

For the Hopf link shown in Fig.~\ref{easyLink}, the perturbations required to alter the nodal topology fall into the range $0.85<|\delta_i^c|<2.75$. 
Comparing this to Eq. (\ref{M_lattice},\ref{linkHamiltonian}), we find that the perturbation required to dismantle the link generally needs to be at least of the same order of magnitude as the hopping integral. Thus, this construction is not only very simple, with hopping terms of only one unit cell, it also features an extraordinary stability to perturbations which is particularly encouraging from an experimental perspective. 


{\it Experimental realization.---} There are several physical settings that have the potential to host knotted non-Hermitian metals, including photonic and acoustic metamaterials, cold atoms \cite{EPrings}, and even strongly disordered and interacting materials \cite{disorderRing,disorderlinesribbons,yoshidapeterskawakmi}. Each of these prominently feature dissipation which is a prerequisite for an effective non-Hermitian description and especially the artificial systems enjoy a high degree of tunability by design. In particular, photonic systems \cite{lujoannopoulossoljacic,ganainymakriskhajavikhanmusslimanirotterchristodoulides}, where the photonic band structure can be directly probed by scattering experiments, are promising due to their high degree of control, especially in the light of the very recent experimental realization of both bulk Fermi arcs \cite{NHarc} in two-dimensional systems and exceptional rings in three dimensions \cite{EPringExp}. 
An especially interesting implementation of this system type is the newly conceived and implemented spectral photonic lattice, where lattice sites are represented by different discrete frequency channels, thus allowing for tunable long range hopping \cite{SpectralPhotonics}.

In the context of experimental realizations of non-Hermitian topology, it should be emphasized that the physical interpretation depends greatly on implementation. In Fermionic many-body systems, the knots are elements of the band structure, whilst in bosonic systems or in the context of classical light, they describe the structure of wave functions or classical waves, yet the fundamental topological features are the same. In this work, we use the language of band theory to make contact with previous works on Hermitian line-node semimetals, though the connection to wave-function topology is as natural.

Regarding the specific ingredients of our model, we note that the required large anti-Hermitian component $d_{I,z}$ in Eq.~(\ref{dvector}) needed in our construction of exceptional knots is straightforwardly implemented via a staggered loss (and/or gain) profile within the unit cell \cite{NHexp,NHexp2}. While asymmetric hopping corresponding to $d_{I,x}$ is more demanding, such terms have also been experimentally realized \cite{asymhop1,asymhop2}. 
Thus, although the realization of the tight-binding models that we propose is certainly demanding, given the required control over hopping parameters with a range of two unit cells, all the basic ingredients for realizing knotted non-Hermitian metals are already available. 
As a precursor of single knots, at least the experimental realization of NH nodal links becomes an immediately experimentally feasible milestone due to our simplified construction (\ref{linkHamiltonian}). There, in contrast to previous work on corresponding Hermitian systems \cite{Nodal-link semimetals,branchCutSemimetals,nodalknotsemimetals}, only nearest neighbor hopping is required, and the resulting tight-binding model is extremely stable against generic perturbations.  

{\it Discussion.---}
In this work we have introduced the notion of {\textit {knotted non-Hermitian metals}} as a class of stable topological phases of matter. Remarkably, despite their apparent complexity, we have shown by explicit construction that non-trivial examples of knotted NH metals can be robustly realized in quite simple tight-binding models. 

To clarify in what sense our findings conceptually go beyond the Hermitian realm, we would like to emphasize two crucial differences to the recently studied nodal knots in Hermitian systems \cite{nodalknotsemimetals}. First, simple parameter counting shows that in Hermitian systems, the occurrence of nodal lines requires fine-tuning or the presence of symmetries, since there only isolated nodal points are generically stable. By contrast, the NH nodal knots discussed in our present work are stable to any small perturbation. Second, the protected open Fermi surfaces occurring in the intriguing form of Seifert surfaces in knotted NH metals are a direct consequence of the fact that we consider knots of exceptional lines rather than ordinary (Hermitian) nodal lines. The non-diagonalizability of such NH knots thus necessarily leads to topological {\textit{metals}}, while nodal knots in symmetry preserving Hermitian systems entail {\textit{semimetals}.

In a broader context, a deep connection between knot theory and topological quantum field theories in (2+1)D was first revealed in a seminal paper by Witten \cite{Witten1989}. Later on, the theory of topological Bloch bands has been viewed as a non-interacting special case of such topological quantum field theories \cite{Qi2008}. Our present analysis of non-Hermitian systems reveals another aspect to the role of knots in the topological classification of matter, where metallic non-Hermitian band structures are directly distinguished by the knot invariants associated with their open Fermi surfaces.

\acknowledgments
{\it Acknowledgments.---} We thank Flore Kunst, Lukas R\o{}dland, and Carsten Timm for useful discussions. 
J.C., M.S. and E.J.B. are supported by the Swedish research council (VR) and the Wallenberg Academy Fellows program of the Knut and Alice Wallenberg Foundation. J.C.B. acknowledges financial support from the German Research Foundation (DFG) through the Collaborative Research Centre SFB 1143 (project-id 247310070).

\end{document}